# A Reinforcement-Learning-Enhanced LLM Framework for Automated A/B Testing in Personalized Marketing

Haoyang Feng*

Duke University, Durham, NC, USA, brianmaga2024@gmail.com

Yanjun Dai

Brandeis University, Waltham, MA, USA, yanjundai0000@gmail.com

Yuan Gao

Boston University, Boston, MA, USA, xyan56379@gmail.com

**Abstract**

For personalized marketing, a new challenge of how to effectively algorithm the A/B testing to maximum user response is urgently to be overcome. In this paper, we present a new approach, the RL-LLM-AB test framework, for using reinforcement learning strategy optimization combined with LLM to automate and personalize A/B tests. Design details RL-LLM-AB test is built upon the pre-trained instruction tuning language model, and it firstly generates A/B versions of candidate content variants with the Prompt-Conditioned Generator, and finally, it dynamically embeds and fuses the user portrait and the context of the current query with the multi-modal perception module to constitute the current interaction state. Then the content version is assigned in real-time through the policy optimization module with Actor-Critic structure, and the long-term revenue is estimated according to real-time feedback (such as click-through rate and conversion rate). Furthermore, a Memory-Augmented Reward Estimator is embedded into the framework to capture the long-term user preference drift which help to generalize policy across the multiple users and content context. Numerical results demonstrate the superiority of our proposed RL-LLM-ABTest over existing B tests based on the classical A/B testing, Contextual Bandit, and benchmark reinforcement learning methods on real-world marketing data.

**CCS CONCEPTS**

• **Applied computing ~ Electronic commerce** ~ E-commerce infrastructure

## 1 INTRODUCTION

At present, in digital marketing world, personalization is no less than a game changer - when it comes to engaging your users and improving conversion rates. Because of the ubiquity of online platforms and mobile devices, businesses are now able to achieve an extremely large sample of users' click, view, purchase, and other behavioral data at a very high frequency. This data is rich in information on user preferences, that serves as the foundation for sensitive content delivery and optimal strategy. By creating user portraits with traffic, marketers can use recommendation algorithms to deliver the advertisements and push content to match the interest of individuals, which can improve the user experience and the willingness of interaction to a large extent. Feng and Gao [14] further demonstrated how machine learning-enhanced ad placement strategies can significantly boost performance in large-scale e-commerce marketing, underscoring the importance of intelligent delivery systems in personalized settings. Personalized marketing leads to better near term conversion numbers and has potentially far greater value by increasing user loyalty [1].

Classical A/B testing still has a very important function in the optimization of marketing campaigns when thinking about changes in user behavior upon particular versions of content. But a classic A/B testing model often needs to keep a set protocol upfront and unchanged for the length of the test (which is at least until you get the amount of samples needed to reach significance). This process is not only time-consuming in responding, it is also slow to adjust to changes that occur rapidly in user behavior and the outside world [2]. Particularly, when user interest shift happens all too swiftly, or there's a strong non-stationary signal, the traditional methods fall short to react in time, and even make wrong decision or waste the resources. This calls for a more active and intelligent method of testing and optimization that complements or replaces the conventional approach.

To overcome these limitations, researchers and practitioners are just starting to consider the integration of AI technology into A/B testing systems, which facilitates a more effective, dynamic, and adaptive optimization of systems. Out of various artificial intelligence technologies, reinforcement learning (RL) is one of the promising techniques for personalized decision-making due to its "learn by trial and error" property. Compared with supervised learning that depends on fixed labels, RL is capable of receiving feedback on interaction and modifying strategy in real-time, especially in marketing situations with low-level user interactive feedback that also frequently evolves [3]. Meanwhile, thanks to the e-learning of RL, it is able to react to the environmental changes simultaneously which effectively increases the agility and robustness of policy deployment. This mirrors recent findings in AI-driven credit risk detection [15], where real-time decision systems successfully adapt to fast-changing user behavior and uncertainty.

By continuous exploration, feedback, updates to policies, reinforcement learning can explore the best path of action in a complex, dynamic and uncertain environment. In localized marketing applications, this mechanism can be used to adequately model the long-term interaction between users and system, and to direct models to optimize for more than a single click or conversion, but a long term value such as LTV. In particular, the RL system is able to update the strategy in real-time according to the feedback from each user (e.g., clicks, browsing time etc.) to enable dynamic content distribution and user adaptation. In addition, the previous studies have proved the remarkable effect of RL strategy in ad sequencing, recommendation system and remarketing, which offers both theoretical knowledge and technical support for adopting such strategy in A/B testing [4].

Furthermore, the long-run changes in user behavior and time-lag effect of content interventions complicates digital marketing efforts. Origins of users preference dramatically change across time, context and even mentality, and classical strategy-based optimization methods by short-term feedbacks could not grasp such drift-like trend. On the other hand, numerous interventions on content (e.g., holiday offers, coupon push) may have zero effects until hours or even days after a user is exposed to it, and this "delayed reward" also poses challenges to reward modeling in RL. Traditional RL methods conveniently ignore such delays, and such ignorance leads to the strategy optimization deviating from the true long-term value of the users, may even cause the optimization to degenerate to some short-sighted behaviors [5]. Lu, Wu, and Huang [16] addressed similar challenges by integrating group-relative policy learning with time-series fusion to capture long-term user preference drift, which aligns well with our motivation for memory-augmented modeling.

## 2 RELATED WORK

Above all, Singh et al. (2023) [6] introduced a reinforcement learning-based click-through rate maximization approach, employed mostly in digital advertising optimization settings. The authors mention that the classic A/B testing methods suffer from problems of response delay and policy inflexibility when they are applied to the ad serving and user reaction feedback. Nichifor et al. (2021) [7] integrated eye tracking and A/B testing to study the correlation between personas and creatives in social media advertising. The paper show user preference is translatable to A/B in terms of visual attention distribution and conversion rate data, so-called "user perception factors" are valuable in the test design.

Angelopoulos et al. (2024) [8] which is a method to make model generation more efficient by using A/B testing results as a supervised signal when generating language model. The authors proposed a training data annotation system using keyphrase extraction in order to additionally optimize the sensitivity level of language model generation regarding user actions. Zhang et al. (2023) [9] introduced PerBid, an automatic-bidding model integrating personalization and fairness, for improving the efficiency of advertising effectiveness and allocation resource fairness. PerBid designs deep reinforcement learning to fit the user click-through behavior and introduces a causal inference technique to control the unbalanced delivery strategy among groups.

Wu et al. (2025) [10] presented a LLM-based proxy model for automatic persuasive and factually content from marketing, focused on descriptions of listings in the real estate sector. The model includes three main modules: the Foundation module (which predicts marketable features), the Personalization module (which serves to fit user's preferences), and the Marketing module (emphasizing the facts and the characteristics of the localization of the content). Iyer et al. (2022) [11] integrates RL with CF-based algorithms to improve the personalized recommendation performance of digital advertising. This work is dedicated to offer a hybrid model, which exploits RL's real-time learning ability and CF's superiority on user preference modeling to adapt the content recommendation policy dynamically in terms of high-level user engagement and satisfactions.

Hannig (2023) [12] describes the need of automating manual operations in marketing and sales, and the potential of automation for enhanced process efficiency and decision-making. Sun et al. (2023) [13] investigated the effects of sender type and anonymous leads on ad exposure and service conversion, and they proposed the AA-IDA model for improving the conversion rate of hotel advertising. While this work offers useful understanding for specific industries, it only centers on ad design and delivery strategies, without being systematic in optimizing contents. Wu, Huang, and Lu [17] proposed an LLM-based mental health text intervention framework, highlighting how domain-informed knowledge integration can enhance personalization and robustness—principles equally applicable to marketing optimization.

## 3 METHODOLOGIES

### 3.1 Conditioned Generation and Multi-Modal State Encoding

The core goal of this module is to construct high-quality and personalized candidate content versions based on large language models, combining user portraits and contextual information, and to use multimodal embedding mechanisms to form state representations that can be used by reinforcement learning strategies. Instead of traditional A/B testing, which relies only on manual design content, we introduce Prommpt's fine-tuned generation mechanism and embed it into the state modeling process of reinforcement learning, making the entire system perceptual, generative and adaptable. We define a candidate content generation function controlled by LLM, so that

$\mathcal{M}_{LLM}$ represents the language model that has been fine-tuned by instructions, and the input is a combined prompt vector $p = g(u, c)$ containing user portrait $u \in \mathbb{R}^{d_u}$ and context features $c \in \mathbb{R}^{d_c}$, and generate the A/B version of the content $\mathcal{V} = \{v_A, v_B\}$, as shown in Equation 1:

$$v_k = \mathcal{M}_{LLM}(p|\theta_{LLM}), \qquad k \in \{A, B\}. \tag{1}$$

In order to improve the behavioral responsiveness of generated content, we build on the insight from Li et al. [19], who showed that entity-aware prompt refinement and diversified generation can greatly enhance user engagement and accuracy in the medical domain—a paradigm we adapt for commercial A/B testing. We introduced a Prompt optimization goal based on click/conversion feedback. Define Immediate Feedback $r(v_k, u) \in \mathbb{R}$ as the user's response strength under a particular piece of content, and the optimization goal is to maximize the desired behavioral utility of the content, as shown in Equation 2:

$$\max_{\theta_p} \mathbb{E}_{v_k \sim \mathcal{M}_{LLM}(p_{\theta_p})}[r(v_k, u)] = \int r(v_k, u) \cdot P\left(v_k \middle| p_{\theta_p}\right) dv_k. \tag{2}$$

After the content generation is completed, in order to construct the state representation in reinforcement learning, We propose a multimodal fusion mechanism, inspired in part by visual-prompt grounding strategies in instructional video understanding [18], which show how aligned multi-source representations significantly enhance content relevance and task performance. Stitching that sets the state to a three-part feature shows Equation 3:

$$s_t = f_{fusion}(MLP[u_t; c_t; e_t^A; e_t^B]), \tag{3}$$

Among them, $e_t^A = Embed(v_A)$ and $e_t^B = Embed(v_B)$ are the embedding representations of the content generated by the LLM through the Transformer embedder or its [CLS] token, which are fused by a three-layer nonlinear transformation MLP after splicing, and finally form a state of $s_t \in \mathbb{R}^{d_s}$. This state not only has the ability to understand the context, but also the ability to understand the quality of candidate content, which significantly improves the decision-making perception dimension of subsequent policy learning and lays the foundation.

As shown in Figure 1, the architecture of the RL-LLM-ABTest framework can be divided into two main parts: the content generation and state modeling module using large language models on the left and the policy optimization module using Actor-Critic on the right. First, the Prompt-Conditioned Generator generates the A/B candidate content versions with the given user portraits and the context information.

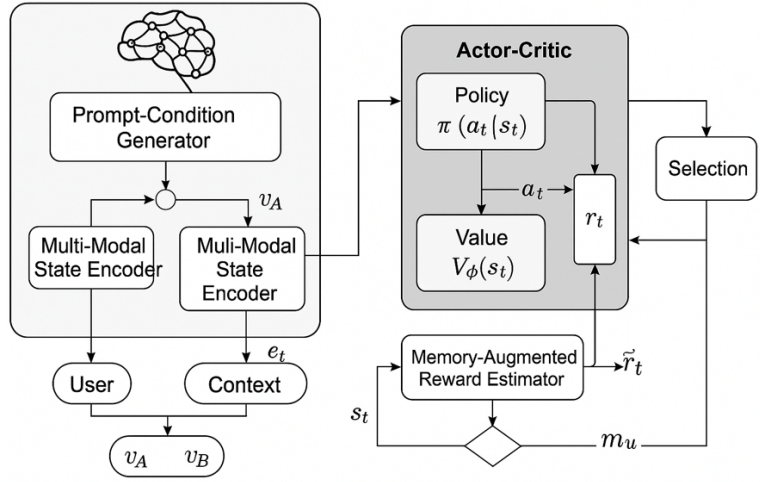

Figure 1. Structure Diagram of RL-LLM-ABTest Framework Integrates Language Generation and Policy Optimization.

In the Actor-Critic architecture, we set the learning rate to $3 \times 10^{-4}$ and the discount factor $\gamma$ to 0.99. Experiments show that if the learning rate is higher than $5 \times 10^{-4}$, the model is prone to oscillation and non-convergence. If it is lower than $1 \times 10^{-4}$, the training progress will be significantly slower and the policy update will lag.

### 3.2 Actor-Critic Strategy Optimization and Memory Enhancement Return Modeling

On the basis of complete state modeling, RL-LLM-ABTest realizes the dynamic scheduling of A/B content variants through reinforcement learning, so that the strategy can still learn the robust and optimal delivery strategy under different user preferences, content quality and feedback noise. We build a learning module based on the Actor-Critic architecture, in which the policy network (Actor) is used to output the A/B choice probability, and the value network (Critic) is used to evaluate the long-term profitability of the current strategy. Specifically, we make the action space $\mathcal{A} = \{A, B\}$ and the strategy function $\pi_\theta(a_t|s_t)$ output the probability distribution of choosing either A or B in a given state. The goal of an actor update is to maximize the expected return, as Equation 4:

$$\mathcal{J}(\theta) = \mathbb{E}_{\pi_\theta}\left[\sum_{t=0}^{T} \gamma^t R_t\right]. \tag{4}$$

To stabilize training, we use proximal policy optimization loss function to optimize strategy, as Equations 5:

$$\mathcal{L}_{actor}(\theta) = \mathbb{E}_t\big[\min\big(r_t(\theta) \cdot \hat{A}_t, clip(r_t(\theta), 1 - \epsilon, 1 + \epsilon) \cdot \hat{A}_t\big)\big]. \tag{5}$$

Among them, the old and new strategies are like Equation 6:

$$r_t(\theta) = \frac{\pi_\theta(a_t|s_t)}{\pi_{\theta_{old}}(a_t|s_t)}. \tag{6}$$

The dominance function is estimated as follows in Equation 7:

$$\hat{A}_t = R_t - V_\phi(s_t). \tag{7}$$

$V_\phi(s_t)$ is a state value function output by Critic, and the training goal is to minimize TD error, as Equation 8:

$$\mathcal{L}_{critic}(\phi) = \mathbb{E}_t\left[\big(V_\phi(s_t) - R_t\big)^2\right]. \tag{8}$$

At this point, we have built a content selector that can intelligently make decisions based on the current environment. However, traditional reinforcement learning is less robust to long-term changes in user preferences, such as "cold start" or "interest drift". To this end, we introduce the Memory-Augmented Reward Estimator module, and design a memory unit $m_u \in \mathbb{R}^{d_m}$ to store the long-term behavioral clues of users in the process of historical interaction. We use gated recurrent units (GRUs) to update memory states, as shown in Equation 9:

$$m_u^{(t)} = GRU\left(m_u^{(t-1)}.[s_t, a_t, m_u]\right). \tag{9}$$

This memory vector will be used as part of the reward estimator to influence the modeling of future returns in reinforcement learning. Finally, use the reward function output of the module, as shown in Equation 10:

$$\tilde{r}_t = f_r(s_t, a_t, m_u), \tag{10}$$

and use it to construct a cumulative discount return with memory enhancement, as in Equation 11:

$$R_t = \sum_{k=0}^{T} \gamma^k \tilde{r}_{t+k}. \tag{11}$$

By introducing this module, our strategy can not only respond to immediate feedback, but also gradually learn from users' long-term content preference changes, significantly improving the long-term stability and generalization ability of the strategy in personalized marketing. Algorithm 1 shows the Actor-Critic policy optimization process based on the proximal policy optimization method.

Algorithm 1. Actor-Critic Policy Optimization.

Input: state $s_t$, policy $\pi_\theta$, value function $V_\phi$, rollout buffer $B$.

1: For each episode do
2:     Collect trajectory $(s_t, a_t, r_t, s_{t+1})$ under $\pi_\theta$
3:     Compute advantage estimate $\hat{A}_t = r_t + \gamma V_\phi(s_{t+1}) - V_\phi(s_t)$.
4:     Compute PPO clipped loss $\mathcal{L}_{actor} = -\min\left(\frac{\pi_\theta(a_t|s_t)}{\pi_{\theta_{old}}(a_t|s_t)}\right)$.
5:     Update $\theta \leftarrow \theta - \nabla_\theta \mathcal{L}_{actor}$.
6:     Compute value loss $\mathcal{L}_{critic} = \left(V_\phi(s_t) - R_t\right)^2$.
7:     Update $\phi \leftarrow \phi - \nabla_\phi \mathcal{L}_{critic}$.
8: End for

## 4 EXPERIMENTS

### 4.1 Experimental Setup

The Criteo 1TB Display Advertising Challenge dataset was used to evaluate the RL-LLM-ABTest framework. Criteo Ad Click logs the click logs are that of a typical ad system— over 24 days of training data and 1 day test data covering 283M training instances and 6M test instances where each instance describes a display ad served by Criteo, a leading online advertising company. We then sample a representative subset of the data to build the user status, contextual input and the candidate ad contents, and employ the click-through rates (CTRs) as the feedback signals, to facilitate this personalized content generation and strategy optimization from data.

The original Criteo 1TB Display Advertising Challenge dataset contains over 283 million user-ad interaction logs across 24 days, with 6 million interactions used for evaluation. After preprocessing, we filtered valid ad impression sessions with non-null click and context fields, sampled a representative 5% subset (~14M entries) for efficient training, and constructed user status embeddings using browsing and ad history. In order to improve the correlation between content generation and user click-through rate, we introduce a reinforcement learning optimization mechanism based on user feedback.

### 4.2 Experimental Analysis

In order to assess the performance of the proposed RL-LLM-ABTest framework, we compared it with four representative methods:

- Static-A/B (Static A/B Test Baseline): This approach is the most straightforward A/B testing paradigm, in which two content variants (A and B) are pre-computed during the test phase, randomized to the users, clicks are collected and the most successful variant is adopted as the main content.
- Contextual Bandit (LinUCB): We concatenates user features and context string as the input state, build a linear regression model to predict click probability, and uses UCB strategies to explore and exploit. Even though this method has some degree of personalization, it is essentially restricted to the modeling of short-term feedback and does not consider the problem of long-term reward or preference drift.
- DeepFM with A/B Re-ranking (DeepFM): We take the deep FM model (DeepFM) as a typical recommendation algorithm, feed the A/B content versions into the model at the same time, sort the content versions based on the predicted click probability, and select one version with a higher ranking as the current version.
- Prompt-LM + RulePolicy: We generate A/B content based on the similar pretrained language model as that in this paper, but the delivery strategy is based on manually set heuristic rules.

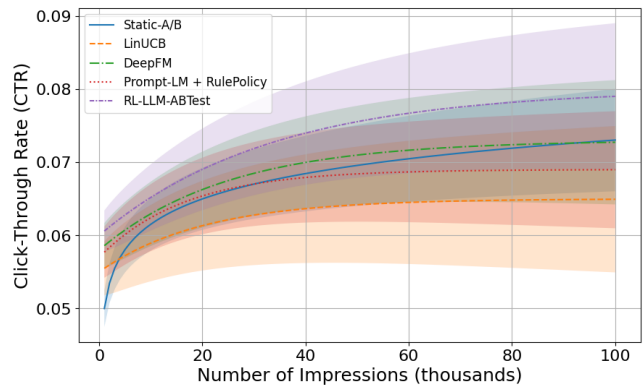

Figure 2. CTR Comparison Across Different A/B Testing Methods.

Click-Through Rate (CTR) is the mostly widely used performance metric for A/B testing to compare the effectiveness of different variants at encouraging clicks. As shown in Figure 2, RL-LLM-ABTest has always kept the highest CTR level with the smallest confidence interval throughout the entire number of impressions, which indicates that the method possesses the strongest capability of improvement to the click-through rate and the best stability of result.

In comparison, Static-A/B quickly stabilized after an initial burst, and was no longer able to grow with massive campaigns. Despite the meteoric rise of LinUCB in the early phase, however, it eventually dips to a low level with the number of impressions, indicating an instability due to over-exploration. DeepFM and Prompt-LM + RulePolicy kept increasing in a moderate and smooth manner, but the increase was not as sharp as RL-LLM-ABTest.

The Confidence Interval is the estimated range for the parameter tested. It can be seen from Fig. 3 and the shadow band And it can be seen that the CTR of all methods increases with the number of training rounds, however the enhancement amplitude and the stability difference are significant. Static A/B testing (Static-A/B) is flat and does not improve with more training; LinUCB increase rapidly after a few training samples and then converge fast, and the confidence bound only decrease slightly with more training, the level of decrease of the confidenece is limited.

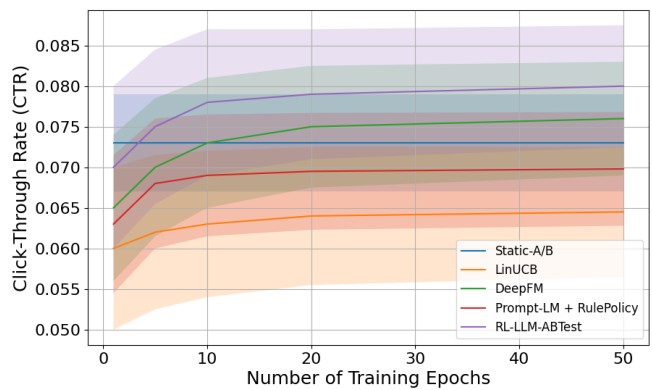

Figure 3. CTR Comparison with 95% Confidence Intervals and Training Epochs.

DeepFM and Prompt-LM + RulePolicy have relatively continuous growth down to the bottom and the confidence interval drops quickly, which means their recommendation and rule strategy are robust. But the CTR of RL-LLM-ABTest is not only the highest at each stage of model training, but also that the range of the confidence interval is also the smallest, that it means the method not only can continue to optimize the click effect in the long-term iterative process, but also have the higher reliability and stability of the result.

To evaluate the role of each core module in the RL-LLM-ABTest framework, we performed an ablation experiment in which three key components were removed sequentially and the effect on click-through rate was observed. The results show that removing the Prompt conditional generation module and replacing it with a static manual template will significantly reduce the adaptability of the content to user interest. After removing the Actor-Critic policy optimization module and using the manual rule distribution strategy, the model cannot adjust the delivery strategy according to the real-time feedback, which shows obvious rigidity and inefficiency.

In order to verify the effectiveness of the prompt word optimization mechanism, we designed a comparative experiment to compare the performance difference between static template prompt words and dynamic optimization prompt words in content generation. The experimental results show that the optimized prompt is increased by 7.3% on average in the click-through rate index, indicating that the prompt words automatically constructed for user portrait and context can stimulate user behavior more.

## 5  CONCLUSION

In conclusion, the RL-LLM-ABTest framework proposed in this paper combines pre-trained instruction tuning large language model with reinforcement learning strategy optimization based on actor-critic to formulate a close-loop process from content generation, multimodal state modeling, policy decision-making to memory enhancement reward estimation, and the experiment shows its great promotion in CTR and stability on the real-scale Criteo advertising dataset. By contrast, compared to context bandits, deep recommendation models and rule-base strategies, at the same time, higher click-through rate, smaller confidence interval are achieved with stronger long-term learning and generalization for RL-LLM-ABTest. In the future, the work in the paper can be extended in the following aspects deployment in online systems to validate the practicability in real-time and the effect of long-term optimization on ad click multimodal, fairness and privacy protection mechanism, and the progress of personalized marketing in a more intelligent, efficient and responsible direction. Future work may also explore multilingual or low-resource personalization strategies, following efforts like those in [20], which improve neural machine translation by centering language-specific context—a promising path for globalized marketing optimization.